# THE LATEST RESULTS OF EXPERIMENT ON THE SEARCH FOR DARK PHOTONS WITH A MULTICATHODE COUNTER.

A. V. Kopylov, I. V. Orekhov and V. V. Petukhov

Institute for Nuclear Research of RAS

117312 Prospect of 60$^{th}$ Anniversary of October Revolution 7A The technique is described of the search for dark photons with a multicathode counter. The aim of experiment is to search for diurnal variations of the effect from conversion of dark photons at the surface of the metallic cathode. The results obtained within 160 days of the measurements are presented.

## 1. INTRODUCTION.

According to our current understanding dark matter makes up about 84% of the mass [1]. We know for sure that dark matter exists but what the nature of dark matter is, it's still a mystery. To solve this enigma, this is a challenge for the present physics. At present there are many experiments aimed at this task but only the upper limits of the model parameters have been obtained so far. One of the possible candidates for this role is dark photon [2 – 6]. We have developed a technique for the search of dark photons with a multicathode counter [7, 8]. This technique enables to find vector direction from diurnal variations of the count rate of single electrons emitted from a metallic cathode of the counter as a result of the conversion of dark photons at the surface of the cathode, if vectors of electric or magnetic fields of dark photons have certain orientation in stellar or solar system. The essential point is that due to rotation of the Earth the annual variations should be symmetrical relative some moment when vector of electric or magnetic field is in the plane of meridian where the detector is situated. What will be the specific curve of variations it depends on the physics of the conversion of dark photon, but the very fact of the symmetry should be present and be independent of specific details. Taking into consideration that we still don't have a robust theory of dark photon we should be ready that real variation curve may turn out to be very surprising for us. Hopefully, one can use as a strong argument in favor of the real detection of the effect from dark photons from the difference in length of the stellar and solar day: 24 h for the solar day and 23 h 56 min for the stellar one [9, 10].

## 2. THE DESCRIPTION OF THE EXPERIMENT.

We use a cylindrical gaseous detector as a proportional counter with an iron cathode 166 mm in diameter and 500 mm long. The counter is filled by a mixture of Ne + CH$_4$ (10%) at pressure

of 0.1 MPa. The detector is placed at the ground floor of a building in Moscow, Troitsk at 55⁰ 45'
N in a shielding cabinet with 30 cm of steel and 10 cm of boronated polyethylene [7, 8]. Steel
shielded a detector from the external gamma-radiation, boronated polyethylene – from thermal
neutrons. Detector is placed horizontally; axes of the detector is in the direction 23⁰ to the one of
North-South. Figure 1 shows the expected diurnal variations if the effect is proportional to $\cos^2 \theta$, here θ – the angle between vector of electric filed and the surface of the counter. This angle
dependence was taken as a most probable expectation that the effect has a maximum when
vector of electric field is normal to the surface of the counter. Because we still don't have a
robust theory of dark photon one can't exclude that the real angular dependence can be different.
If in experiment the effect from dark photons is detected one can study this in detail. The
expected curves of diurnal variations depend also on the orientation of the detector (vertical,
horizontal, the axis along the North-South or East-West direction), on the geographical latitude
of the place where detector is situated, from the quality of the inner surface of the cathode
(mirror, matt) [11, 13].

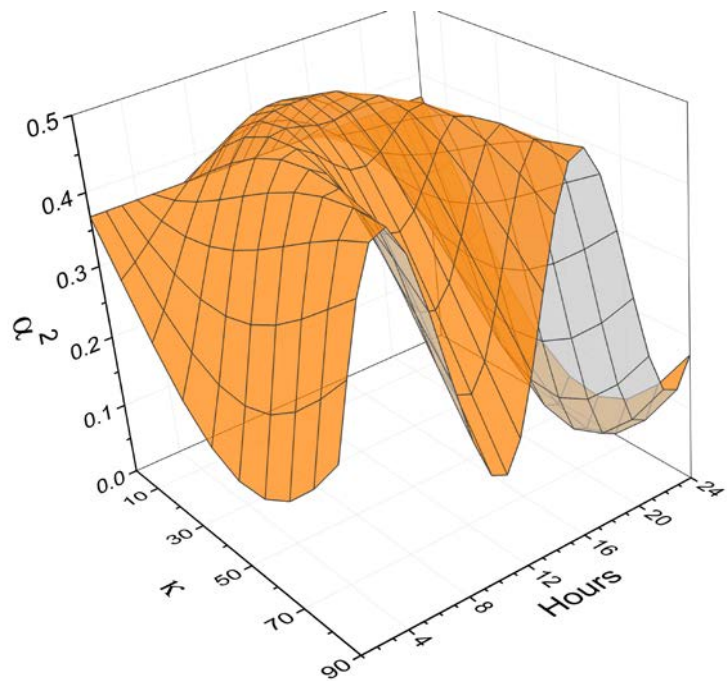

Figure 1. The diurnal variations calculated for the rotation of the Earth. Here $\alpha^2 = \langle \cos^2\theta \rangle$ averaged
through the surface of the cathode, κ – the angle between vector of E-field and the axis of the Earth.

Figure 2 shows slices of diurnal variations made at 30 and 90 degrees of the angle between
vector of E-field and the axes of the Earth. Here for the time 00-00 it was taken a moment when
vector of E-field is in the plane of the meridian where the detector is placed, i.e. meridian of
Moscow where the experiment is conducted. The measurements have been conducted on 24 h basis, 1 TB of data have been collected during a day. The voltage on the output of charge sensitive preamplifier has been submitted to the input of ADC board and digitized with time resolution 100 ns. The data treatment has been conducted in off-line. The pulses with amplitude from 3 to 50 mV has been chosen as candidates for "true" pulses from single electrons. The shapes of the pulses have been compared with the reference pulse obtained during calibration and pulses with shapes of irregular kind have been discarded.

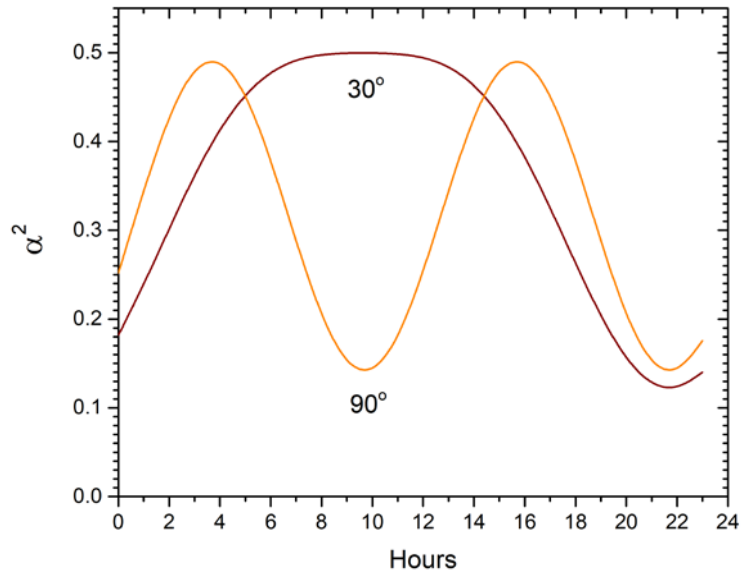

Figure 2. The slices from Fig. 1 at $\kappa = 30^\circ$ and $90^\circ$.

The data have been obtained for 160 days of measurements in total. The measurements have been interrupted for calibration and for control measurements in configuration 2 where background has been measured. Then the measured count rates have been grouped in 2h intervals in stellar and solar time. The corresponding temporal distributions have been drawn from these data. Figure 3 shows the ones for sideral time and Fig. 4 – for terrestrial time. The distributions have been drawn for total time 160 days and 1645 points in 2h interval each point.

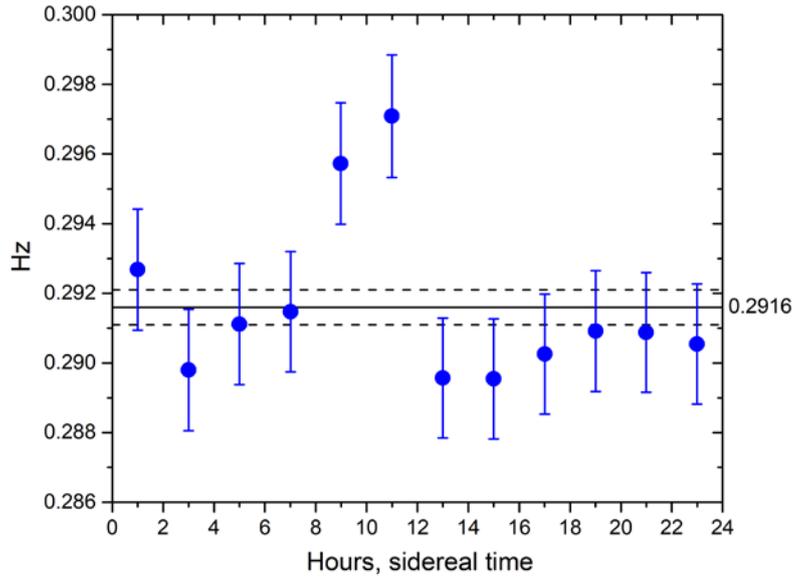

Figure 3. The temporal distribution in sidereal time.

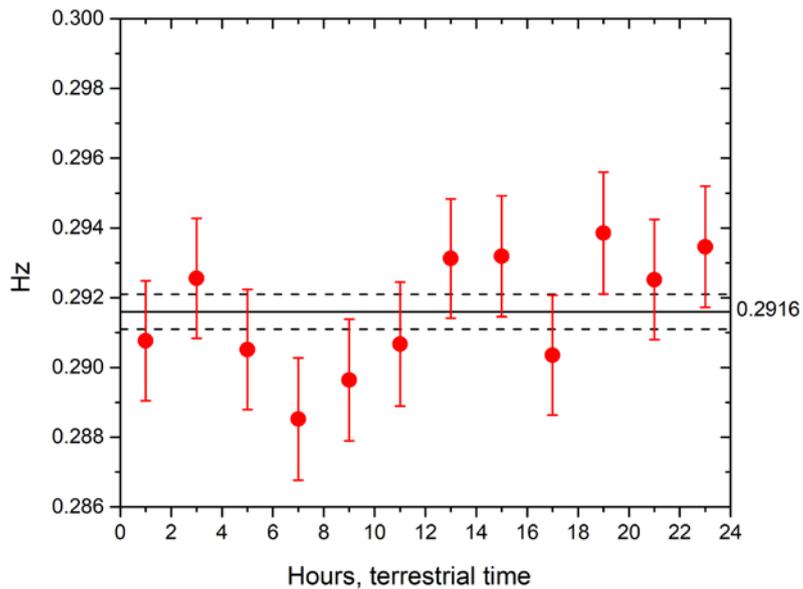

Figure 4. The temporal distribution in terrestrial time.

One can see from Fig. 3 that in time interval from 8-00 till 12-00 we have substantial excess of count rate in comparison with the average one. We don't see a similar excess in Fig. 4. For more detailed analysis of this effect, we have built the distributions for shorter intervals of time. The first and second ones composed 548 points, the third one – 549 points. Figure 5 shows the corresponding distributions for siderad time, Fig 6 – for terrestrial time. One can see at Fig. 5 that systematic excess of count rates is observed in time interval from 8-00 till 12-00 by all three distributions. For terrestrial time we don't observe any similar excess as one can see from Fig. 6.

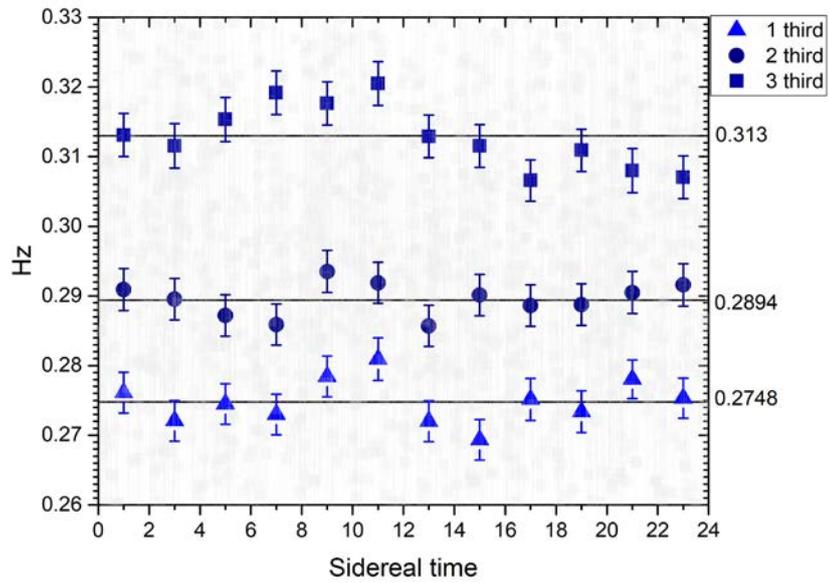

Figure 5. The temporal distribution in sidereal time.

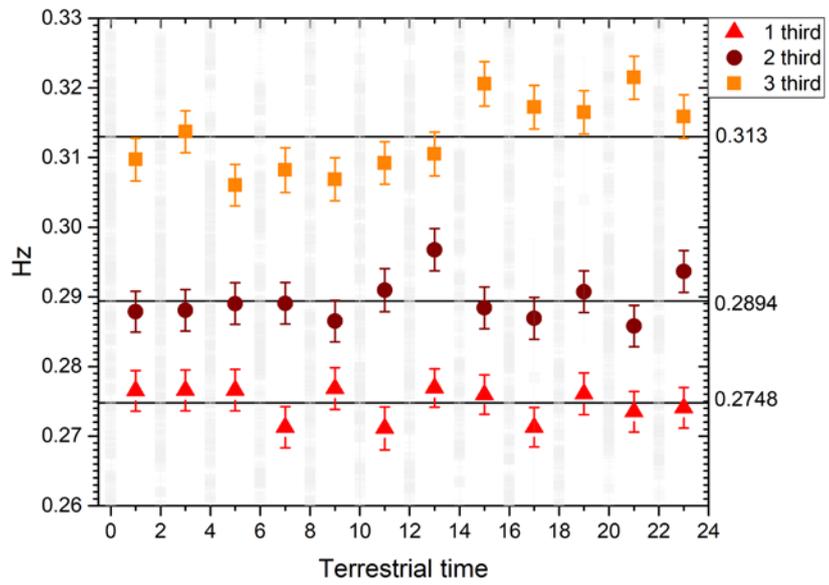

Figure 6. The temporal distribution in terrestrial time.

## 3. CONCLUSIONS.

We have measured the count rates of single electrons with a multi-cathode counter. By the results obtained during 160 days of measurements (1645 points) we have drawn temporal distributions for sideral time and terrestrial time for each 2 h intervals of three short periods of approximately equal times. For the distributions in sideral time we observe substantial excess of count rates in comparison with the average one in the time interval from 8-00 till 12-00 sideral time. The estimated confidence level of this excess is on the level of 3.5 σ. For the terrestrial time we don't observe any such effect. We continue measurements with the aim to increase the confidence level. We consider this result as a preliminary one.

Acknowledgements: We appreciate very much the substantial support from the Ministry of Science and Higher Education of Russia Federation within "The Instrument Base Renewal Program" in the framework of the State program "Science".